\documentclass[letter,11pt]{article}
\usepackage{float}
\usepackage{graphicx, amssymb, wrapfig,setspace,multicol,natbib,hyperref,setspace}
\def\ltsima{$\; \buildrel < \over \sim \;$}
\def\simlt{\lower.5ex\hbox{\ltsima}}
\def\gtsima{$\; \buildrel > \over \sim \;$}
\def\simgt{\lower.5ex\hbox{\gtsima}}
\newcommand{\uJy}{$\mu$Jy}

\newcommand{\lsun}{L$_{\odot}$}
\newcommand{\msun}{M$_{\odot}$}

\begin{document}

\begin{picture}(30,100)
\vspace{-1.5cm} 
\put(23,145){  {\sc \large \kern0.8cm Imaging Cold Gas to 1\,kpc scales in high-redshift galaxies}}
\put(20,132){ \large \kern0.88cm {\sc with the ngVLA}}

\end{picture}
\vspace{-3.8cm}
\hrule width \textwidth
\vspace{0.2cm}

\vspace{2mm}
\begin{figure}
  \includegraphics[height=1.5cm]{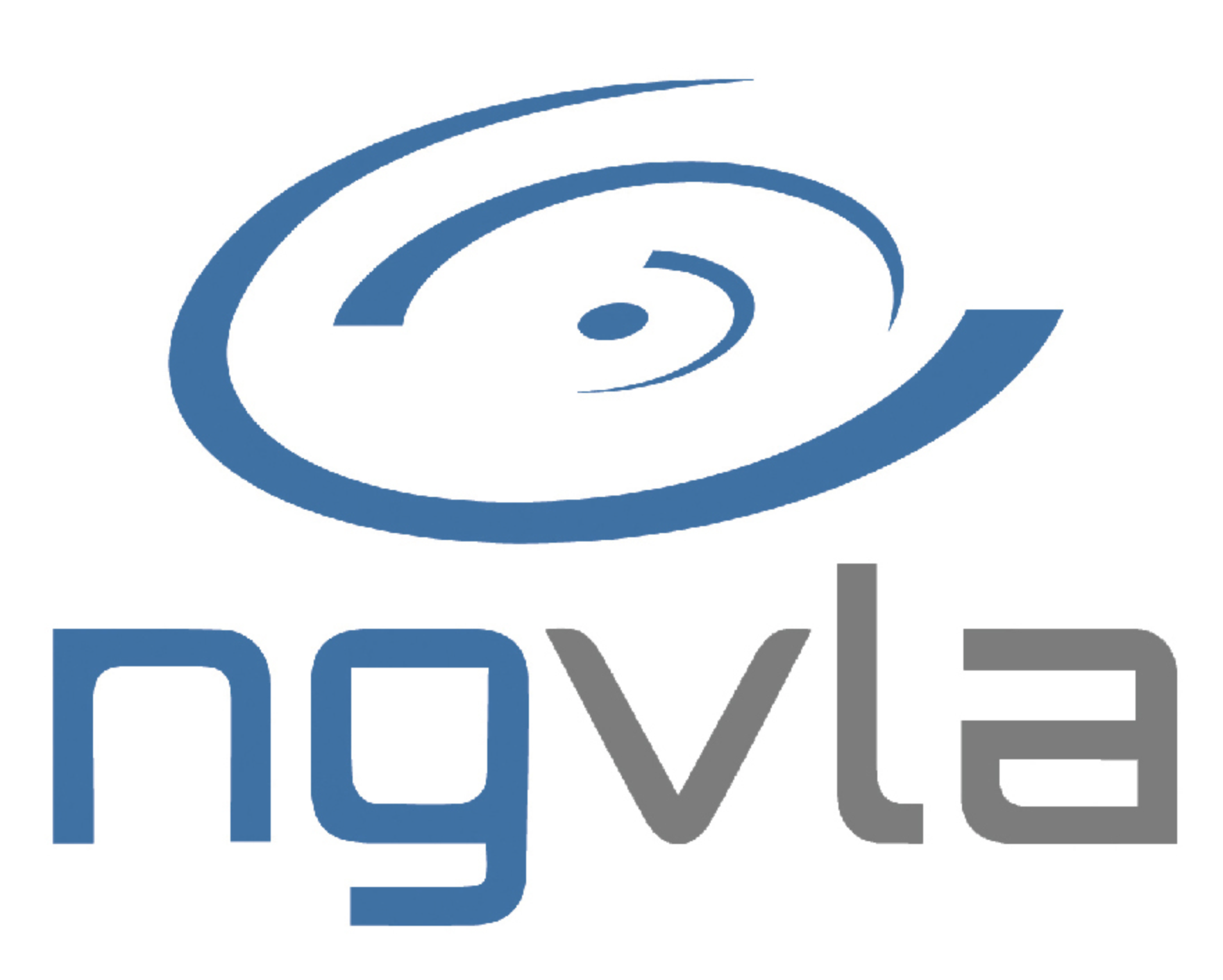}

\end{figure}

\vspace{-0.3cm}

\begin{flushright}
\today
\end{flushright}


\noindent Caitlin M. Casey$^{1}$, 
Desika Narayanan$^{2,3}$,
Chris Carilli$^{4}$, 
Jaclyn Champagne$^{1}$,
Chao-Ling Hung$^{1}$,
Romeel Dav\'{e}$^{5}$,
Roberto Decarli$^{6}$,
Eric J. Murphy$^{7}$,
Gergo Popping$^{8}$,
Dominik Riechers$^{9}$,
Rachel S. Somerville$^{10,11}$,
Fabian Walter$^{8}$

\begin{spacing}{0.7}
{\footnotesize
\noindent $^{1}$ Department of Astronomy, The University of Texas at Austin, 2515 Speedway Blvd, Austin, TX 78712, USA\\
\noindent $^{2}$ Department of Astronomy, University of Florida, 211 Bryant Space Science Center, P.O. Box 112055, Gainesville, FL 32611, USA\\
\noindent $^{3}$ Cosmic Dawn Center (DAWN), Niels Bohr Institute, University of Copenhagen, DK-2100 Copenhagen $\O$; DTU-Space, Technical University of Denmark, DK-2800 Kgs. Lyngby\\
\noindent $^{4}$ National Radio Astronomy Observatory, 1003 Lopezville Rd, Socorro, NM 87801, USA\\
\noindent $^{5}$ Institute for Astronomy, University of Edinburgh, Royal Observatory, Blackford Hill, Edinburgh EH9 3HJ, UK\\
\noindent $^{6}$ INAF - Osservatorio di Astrofisica e Scienza dello Spazio di Bologna, via Cobetti 93/3, 40129 Bologna, Italy\\
\noindent $^{7}$ National Radio Astronomy Observatory, 1180 Boxwood Estate Rd, Charlottesville, VA 22903, USA\\
\noindent $^{8}$ Max Planck Institute for Astronomy, K\"{o}nigstuhl 17, 69117 Heidelberg, Germany\\
\noindent $^{9}$ Department of Astronomy, Cornell University, 616A Space Science Building, Ithaca, NY 14850, USA\\
\noindent $^{10}$ Department of Physics \&\ Astronomy, Rutgers University, 136 Frelinghuysen Rd, Piscataway, NJ 08854, USA\\
\noindent $^{11}$ Center for Computational Astrophysics, 162 5th Ave, New York, NY 10010, USA\\
}
\end{spacing}

\begin{center}
\begin{minipage}{12.5cm}
{\bf \large Abstract} The next generation Very Large Array (ngVLA)
will revolutionize our understanding of the distant Universe via the
detection of cold molecular gas in the first galaxies.  Its impact on
studies of galaxy characterization via detailed gas dynamics will
provide crucial insight on dominant physical drivers for
star-formation in high redshift galaxies, including the exchange of
gas from scales of the circumgalactic medium down to resolved clouds
on mass scales of $\sim$10$^{5}$\,M$_\odot$.  In this study, we employ
a series of high-resolution, cosmological, hydrodynamic zoom
simulations from the {\sc Mufasa} simulation suite and a CASA simulator to
generate mock ngVLA observations of a $z\sim4.5$ gas-rich 
star-forming galaxy.  Using the {\sc Despotic} radiative transfer code that
encompasses simultaneous thermal, chemical, and statistical equilibrium in
calculating the molecular and atomic level populations, we generate
parallel mock observations of low-J to high-J transitions of CO from
ALMA for comparison.  We find that observations of CO(1-0) are
especially important for tracing the systemic redshift of the galaxy
and the total mass of the well-shielded molecular gas reservoir, while even CO(2-1)
can predominantly trace denser gas regions distinct from CO(1-0). The
factor of 100 times improvement in mapping speed for the ngVLA beyond
the Jansky VLA and the proposed ALMA Band 1 will make these detailed,
high-resolution imaging and kinematic studies of CO(1-0) routine at
$z\sim2-5$.
\end{minipage}
\end{center}

\vspace{1mm}



\section{Introduction}

Cold molecular gas is the fuel for cosmic star-formation
\citep{solomon05a,carilli13a}.  This gas is most commonly probed
through observations of CO, the most abundant molecule after molecular
hydrogen, and a molecule which is easily excitable at low
temperatures.  Resolved CO studies of distant galaxies can not only
provide the basic measurement of a galaxy's total molecular gas mass,
but one can also extract detailed kinematics of that cold gas
reservoir.  The kinematics give direct constraints on the internal
dynamics of galaxies out to high-$z$ and insight into the physics of
the interstellar medium
\citep[e.g.][]{engel10a,bothwell10a,ivison12a}: whether or not they
are dispersion dominated, agitated by a dynamical interaction, or
rotating in a disk.  Ideally, these kinematic tracers would be
available for a large sample of early Universe galaxies to address
important questions in galaxy formation and evolution, like assessing
the relative role of major mergers in driving the growth of massive
galaxies over cosmic time.

It is not immediately clear that all tracers of molecular gas in
distant galaxies would lead to the same observational conclusions
about galaxy dynamics, as each traces fundamentally different physical
states of the gas.  For example, the lowest ground state transition of
CO, CO(1-0), should trace the most diffuse and massive gas reservoir
in galaxies, while higher-J transitions probe sequentially denser
environments.  Very few galaxies in the early Universe have resolved
observations from {\it multiple} molecular gas tracers, because the
time required to spatially map each transition is currently
prohibitive for large samples of galaxies, even on the most sensitive
interferometers like ALMA.   And
yet, there is some indication that different tracers do point to
fundamentally different gas reservoirs, that could lead to differences
of interpretation of their dynamics \citep{bothwell13c,hodge12a}.


In this study, we investigate quantitative differences in observations
of different CO transitions in a hydrodynamic simulation of a early
Universe massive galaxy.  Based on a direct comparison between the
inferred results from our mock observations and the cosmological
simulations, we investigate the capabilities of ALMA and the proposed
next generation Very Large Array \citep[ngVLA][]{carilli15a} to
constrain the mode of star formation, dynamical mass, and molecular
gas kinematics in individual high-redshift galaxies \citep[see][for a
  broader discussion of ngVLA's use in high-$z$ extragalactic
  astrophysics]{casey15b}.  This work highlights the complexity in
inferring these important quantities in early Universe galaxies, and
the need for the next generation facilities.

\section{Hydrodynamic Zoom-In Simulation Case Study}

\begin{figure}
\centering
\includegraphics[width=0.5\textwidth]{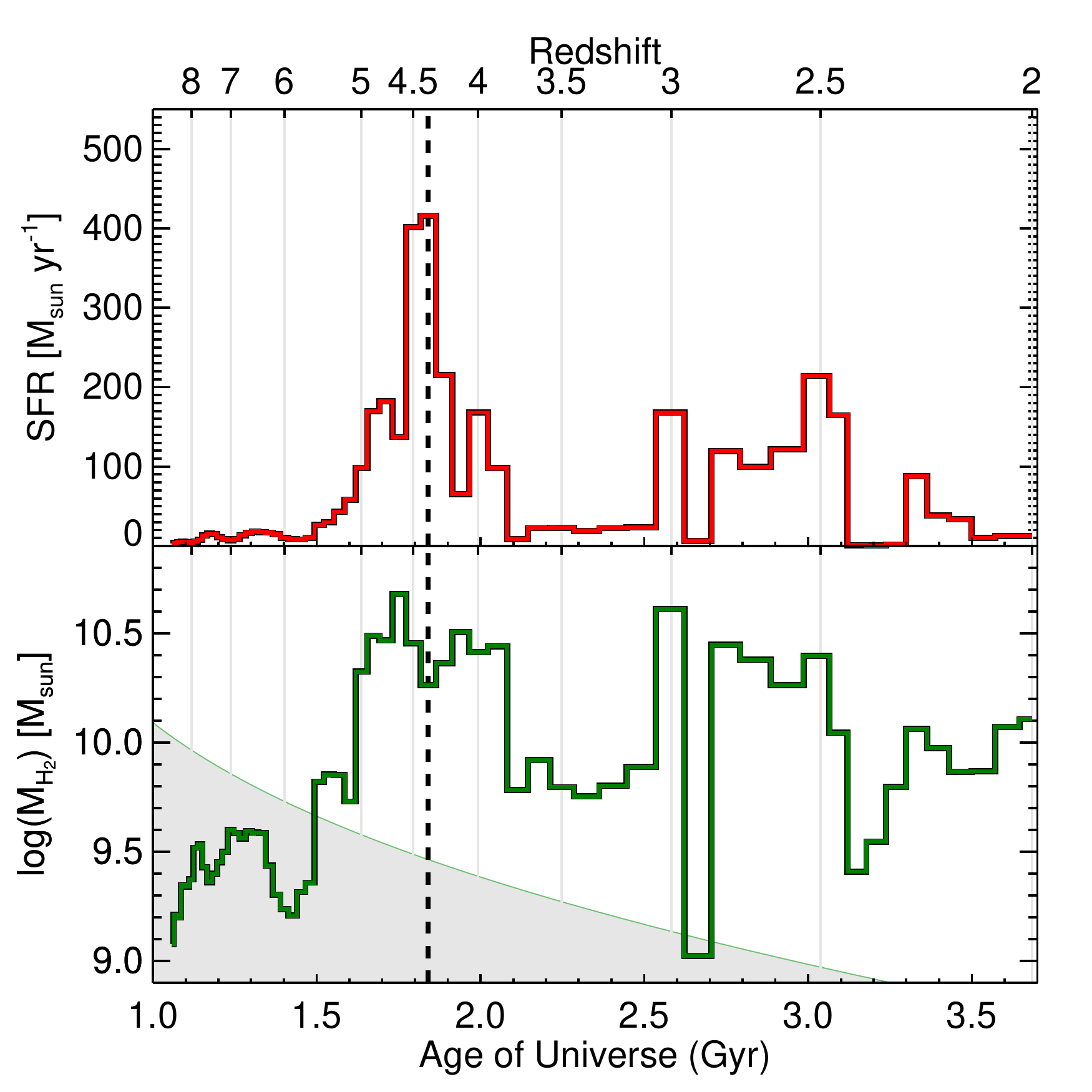}
\caption{The star-formation history (top) and molecular hydrogen mass
  (bottom) of the evolving massive galaxy in formation.  The zoom-in
  simulation is run from high-redshift to $z=2$, at which point it has
  a stellar mass of 2$\times$10$^{11}$\,\msun.  The system undergoes
  two major star-forming events, and we analyze the snapshot where the
  SFR is maximized near 400\,\msun/yr at $z=4.37$ (marked with a
  vertical dashed line).  The thin green line represents the 5$\sigma$
  detection limit for the ngVLA in one hour; to resolve the kinematics
  of the galaxy across $\sim$10 beams, 10$\times$ the observing time
  is required to reach adequate analysis depth.}
\label{fig1}
\end{figure}

We employ a series of high-resolution, cosmological, hydrodynamic zoom
simulations \citep{narayanan18a,narayanan18b} from the {\sc Mufasa}
simulation suite \citep{dave16a,dave17a} and a CASA simulator to
generate mock ngVLA observations.  Using the {\sc Despotic} radiative
transfer code \citep{krumholz14b} that encompasses simultaneous
thermal, chemical and statistical equilibrium in calculating the
molecular and atomic level populations \citep{narayanan17a}, we
generate parallel mock observations of low-J to high-J transitions of
CO.  For this study we take a single snapshot of one zoom simulation.
The galaxy being formed is an extremely massive
$M_\star>4\times10^{11}$\,\msun\ at $z=0$; the zoom in itself is only
run from high-$z$ down to $z=2$, at which point the stellar mass is
already 2$\times$10$^{11}$\,\msun, which is $\approx M_\star$ at
$z=2$.  The system undergoes two `starbursting' episodes during its
history at $z\sim4.5$ and $z\sim2$; the first is driven by the infall
of several small galaxies (i.e. minor mergers) while the later burst
is driven by a major merger.  We have chosen to focus our study on the
earlier $z\sim4.5$ snapshot, which represents the strongest starburst
episode of the galaxy's history.  The galaxy's star formation history
and evolving molecular gas mass is shown in Figure~\ref{fig1}.  At the
sampled snapshot, the galaxy's total stellar mass is
7$\times$10$^{10}$\,\msun, its mass of molecular hydrogen is
2$\times$10$^{10}$\,\msun, and its 50\,Myr-averaged SFR is
415\,\msun\,yr$^{-1}$.  The simulated galaxy's gas kinematic structure
is inherently complex owing to the minor interactions, though
morphologically the galaxy presents as a stable disk \citep[consistent
  with what is found observationally for similar galaxies at $z>2$ in
  dust continuum,][]{hodge16a}.

\begin{figure}
\centering
\includegraphics[width=0.7\textwidth]{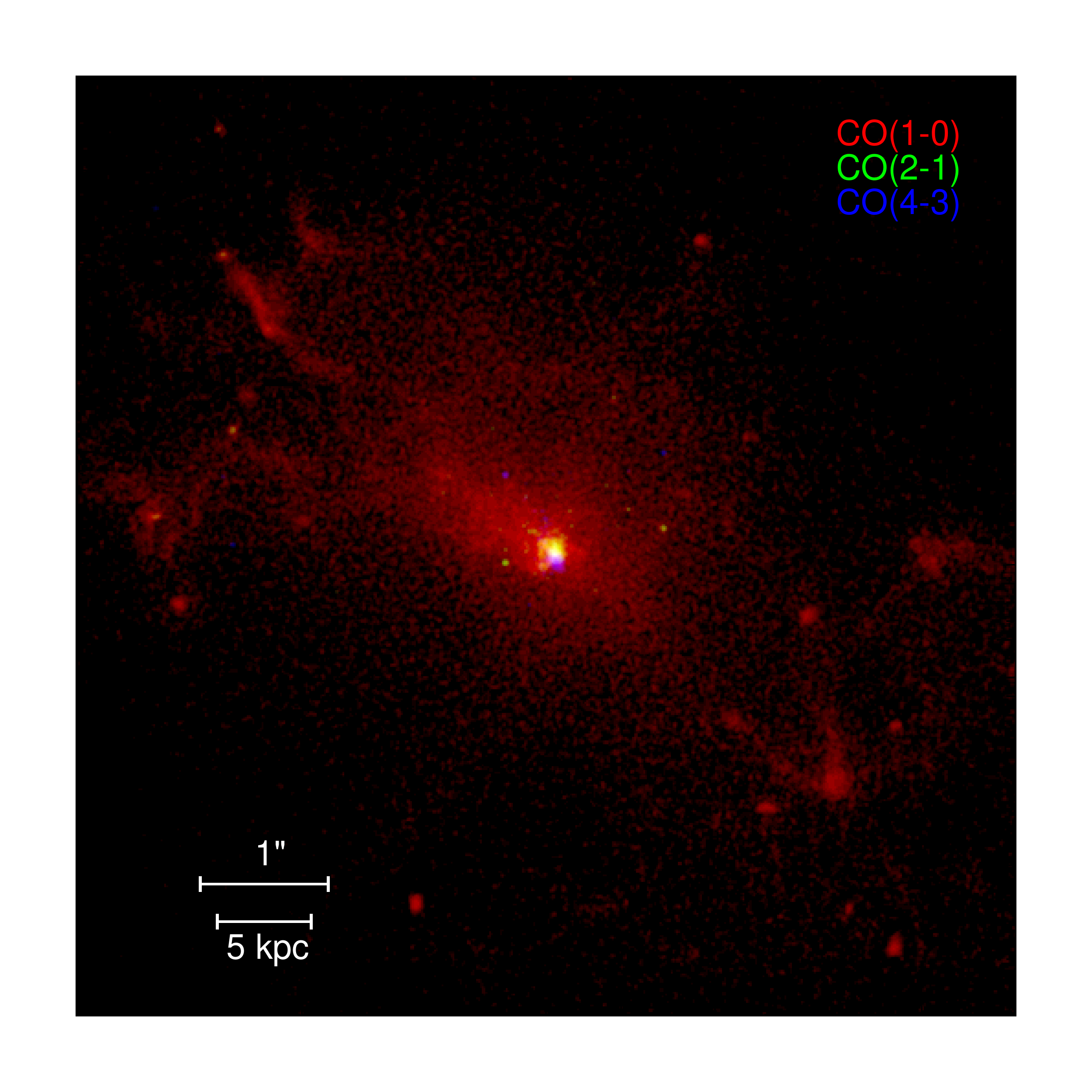}
\caption{A false-color image of molecular gas emission in
  the $z=4.37$ snapshot of the massive galaxy in formation.  CO(1-0)
  emission is traced in red, CO(2-1) in green and CO(4-3) in blue.
  The luminosity scale is weighted to match the peak line intensity at
  the galaxy's core; so while the higher-J transitions are
  intrinsically brighter lines, this scaling shows the physical scale
  over which each line is emitting.  The disk of molecular gas
  extending out beyond $\sim$8\,kpc from the core is only prominent in
  CO(1-0) emission.}
\label{fig2}
\end{figure}

\begin{figure}
\centering
\includegraphics[width=0.45\textwidth]{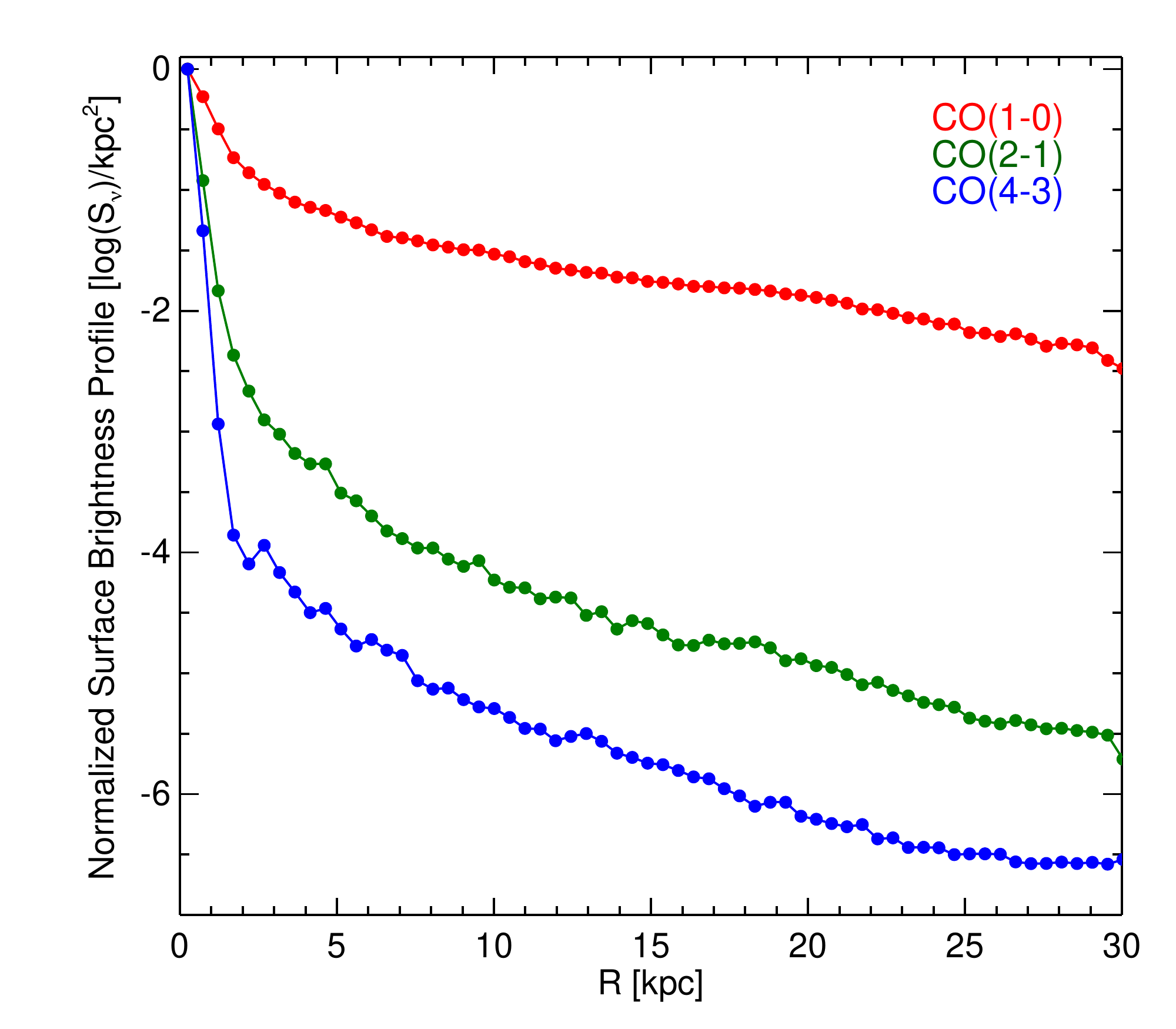}
\caption{ The intrinsic surface brightness profiles of CO(1-0),
  CO(2-1) and CO(4-3) for the simulated galaxy shown in
  Figure~\ref{fig2}, differing by 3-4 orders of magnitude at large
  galactocentric radii.}
\label{fig3}
\end{figure}

We note that this source sampled at this particular time is expected
to be among the brighter galaxies detectable with the ngVLA.
Figure~\ref{fig1} shows the effective 5$\sigma$ detection limit as a
function of redshift for a 1 hour ngVLA observation on CO(1-0).  This
limit is determined by working backward from a target RMS of
10\uJy/beam per 2\,MHz channel in 1 hour of observations, an assumed
average 200\,km/s line width, and a CO-to-H$_{2}$ conversion factor of
$\alpha_{CO}=1$\,M$_{\odot}$\,(K\,km/s\,pc$^{2}$)$^{-1}$.  In other
words, this is the limit at which the total integrated CO(1-0) line
would have a 5$\sigma$ significance if it were 200\,km/s wide.  The
galaxy snapshot studied here would be detectable with a total SNR of
$\sim$32 in 1 hour with the ngVLA.  In contrast, GN20, at a similar
luminosity and redshift, at $z=4.05$ required a total of 120\,hours
with the current VLA.
If, as in this study, one wishes to spatially resolve the CO line
across $\sim$10 beams, 10 hours of integration are required to reach
gas masses at this limit.  If this signal needs to be further resolved
spectrally into $N$ spectral bins (where $N$ has a minimum value of
$\sim$4 to kinematically resolve rotation), the RMS per channel goes
up as $\sqrt{N}$ and the required observing time scales with $N$.

For this particular galaxy snapshot, the integrated CO luminosity is
2$\times$10$^{10}$\,\lsun.  Distributed across $\sim$10 beams and
$\sim$7 spectral bins, the average SNR per beam is anticipated to be
$\sim$10 in a 1 hour ngVLA integration of CO(1-0).  The required
observing time would be approximately 30 minutes for the
$\sim$4$\times$ brighter CO(2-1) transition.  The same system as
observed with ALMA in CO(4-3) at matched SNR per spatial and spectral
bin would require $\sim$12\,hours of on-source integration.

Figure~\ref{fig2} shows a false-color image tracing the molecular gas
luminosity of the simulated galaxy itself, where CO(1-0) emission is
shown in red, relative to CO(2-1) in green and CO(4-3) in
blue\footnote{Note that at this redshift, CO(3-2) is not observable as
  its frequency falls in an opaque atmospheric frequency range.}.  The
entire gas reservoir emits somewhat uniformly in CO(1-0), indicative
of diffuse gas in the galaxy's interstellar and circumgalactic
mediums, rendering this tri-color image largely red over most of the
galaxy's observed surface area.  The radial surface brightness
profiles of CO emission is shown in Figure~\ref{fig3}; the contrast of
peak emission at the core to diffuse in the outskirts is only a factor
of 10-100 in CO(1-0) while it is $>$10$^{4-6}$ for higher-J CO lines,
including CO(2-1).

\begin{figure}
\includegraphics[width=\textwidth]{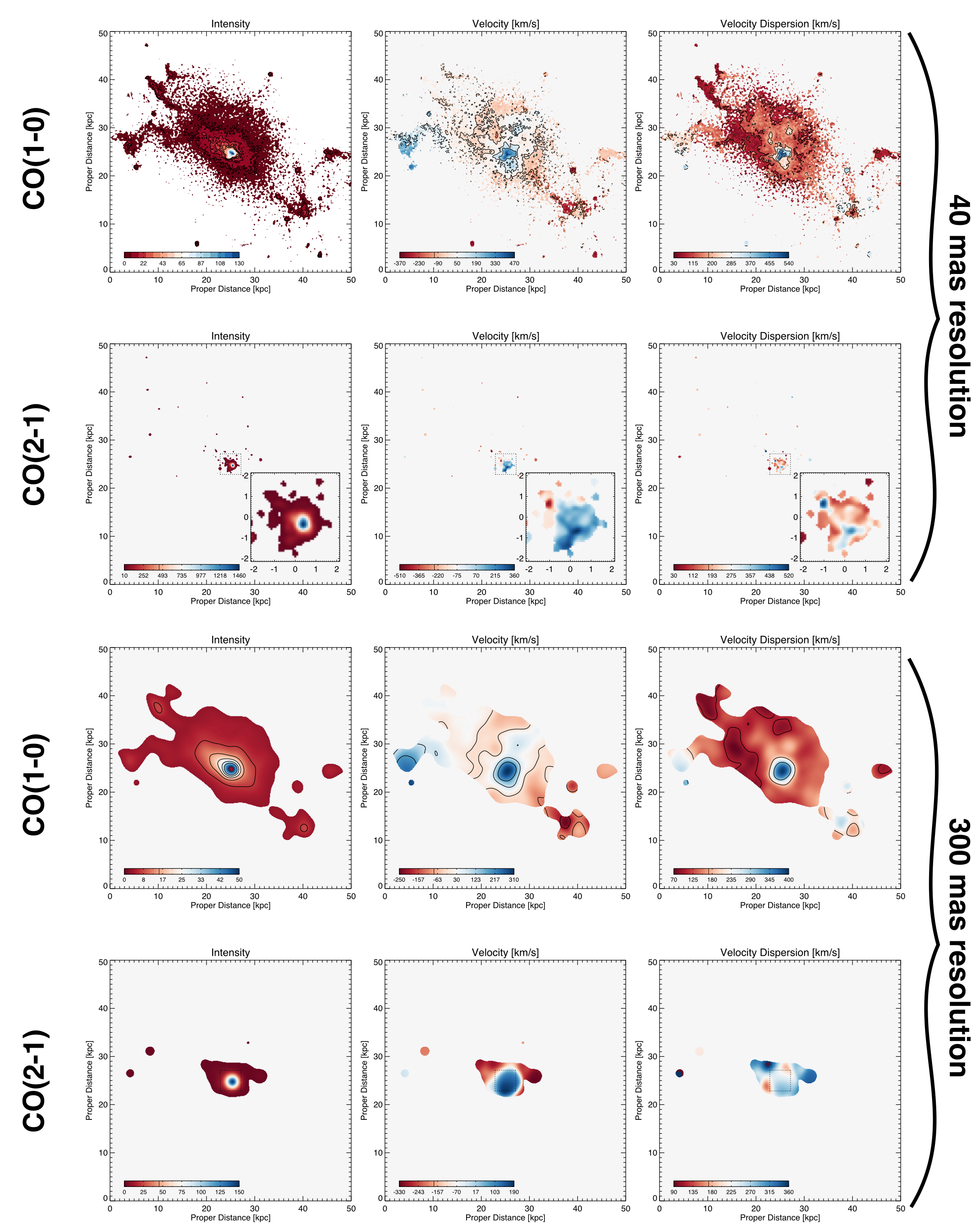}
\caption{CO intensity, velocity, and velocity dispersion diagrams for
  the galaxy snapshot under analysis, in both CO(1-0), first and third
  rows, and CO(2-1), second and fourth rows.  The first two rows show
  the intrinsic kinematic structure of the simulated galaxy with
  40\,mas resolution, corresponding to physical scales of 270\,pc.  At
  bottom, the same diagrams are shown at 300\,mas resolution,
  physically 2\,kpc resolution. }
\label{fig4}
\end{figure}

The kinematic breakdown of the gas in this galaxy is shown at two
different angular resolutions in both CO(1-0) and CO(2-1) in
Figure~\ref{fig4}.  The simulated kinematic breakdown in CO(4-3) is
very similar to the CO(2-1) profile.  All diagrams are limited to a
dynamic range of $\sim$150 from peak intensity, which not only mimics
realistic user-limited depths of future datasets, but avoids complex
calibration needed for high dynamic range interferometric datasets.
In other words, even though there is an extended component of emission
in CO(2-1) and higher J transitions, they are unlikely to be detected
given the significant surface brightness discrepancy between the core
and extended disk, as shown in Figure~\ref{fig3}.  The first two rows
show the gas profile at 40\,mas resolution, equivalent to a physical
resolution scale of 270\,pc at this redshift.  Without any simulated
noise, these diagrams represent the intrinsic kinematic structure of
the simulated galaxy.  Certainly the complexity of the gas reservoir
is quite clear: this system is not a smoothly rotating disk, but
rather a turbulent, gas-rich, dispersion dominated system.  The
difference between the kinematic profiles in CO(1-0) vs CO(2-1) is
striking.  The compact CO(2-1), which would only be observable on
spatial scales $<$1$''$, probes the galaxy's core on
$\sim$2-3\,kpc scales.  Though one would conclude that this galaxy is
dominated by gas turbulence from both CO(1-0) and CO(2-1) kinematics,
only the CO(1-0) data shed light on the more diffuse gas on large
physical scales. 

\begin{figure}
\centering
\includegraphics[width=0.69\textwidth]{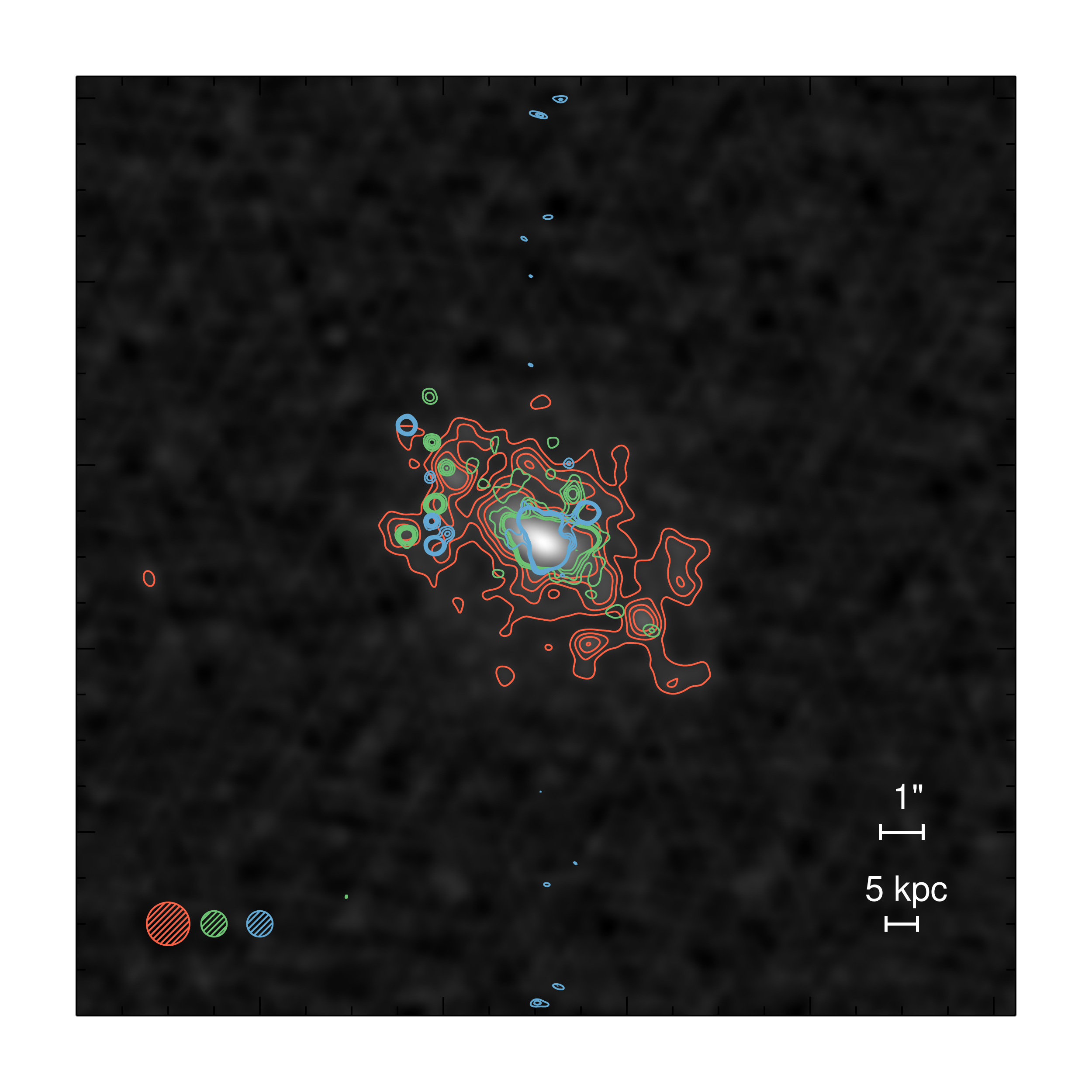}\\
\includegraphics[width=0.49\textwidth]{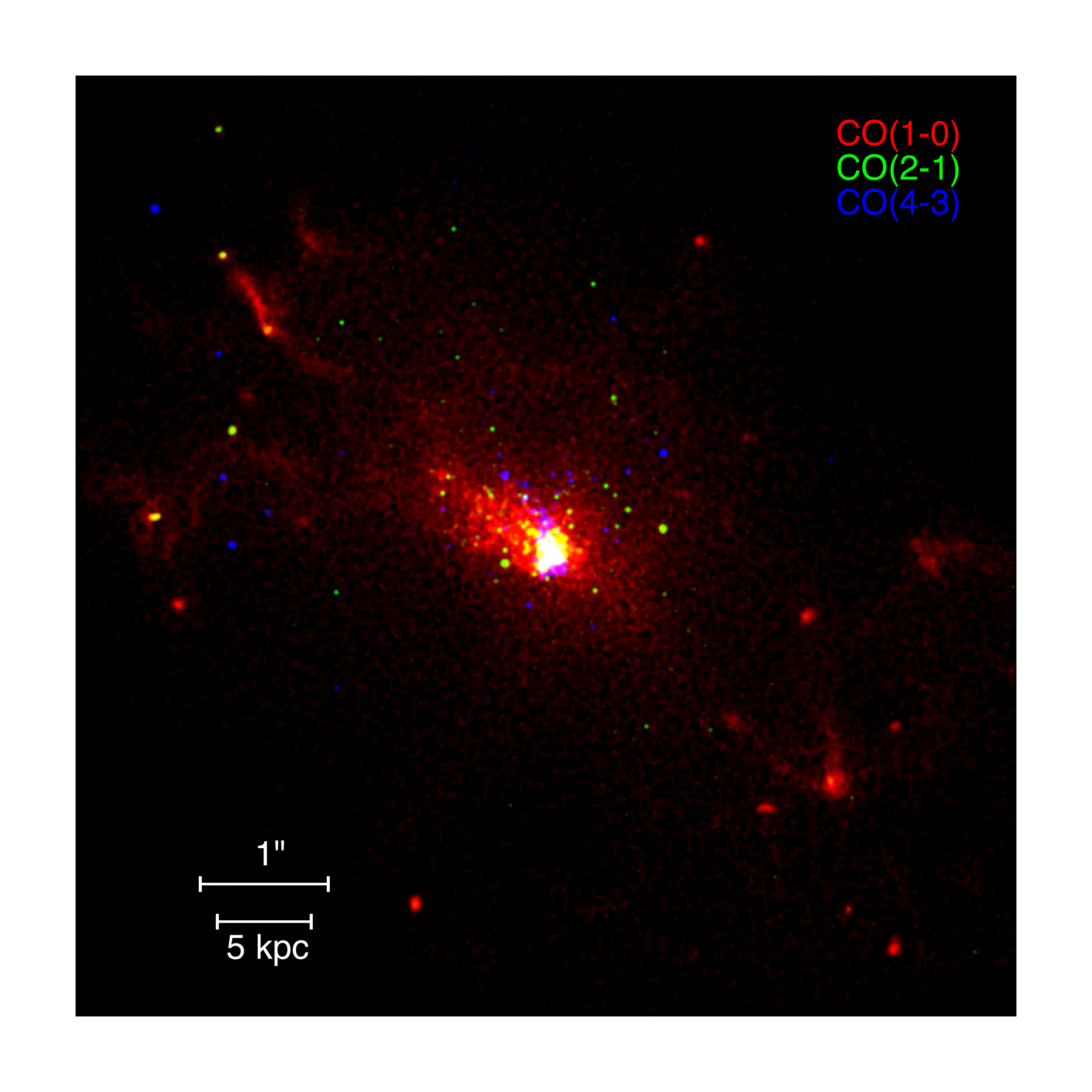}
\includegraphics[width=0.49\textwidth]{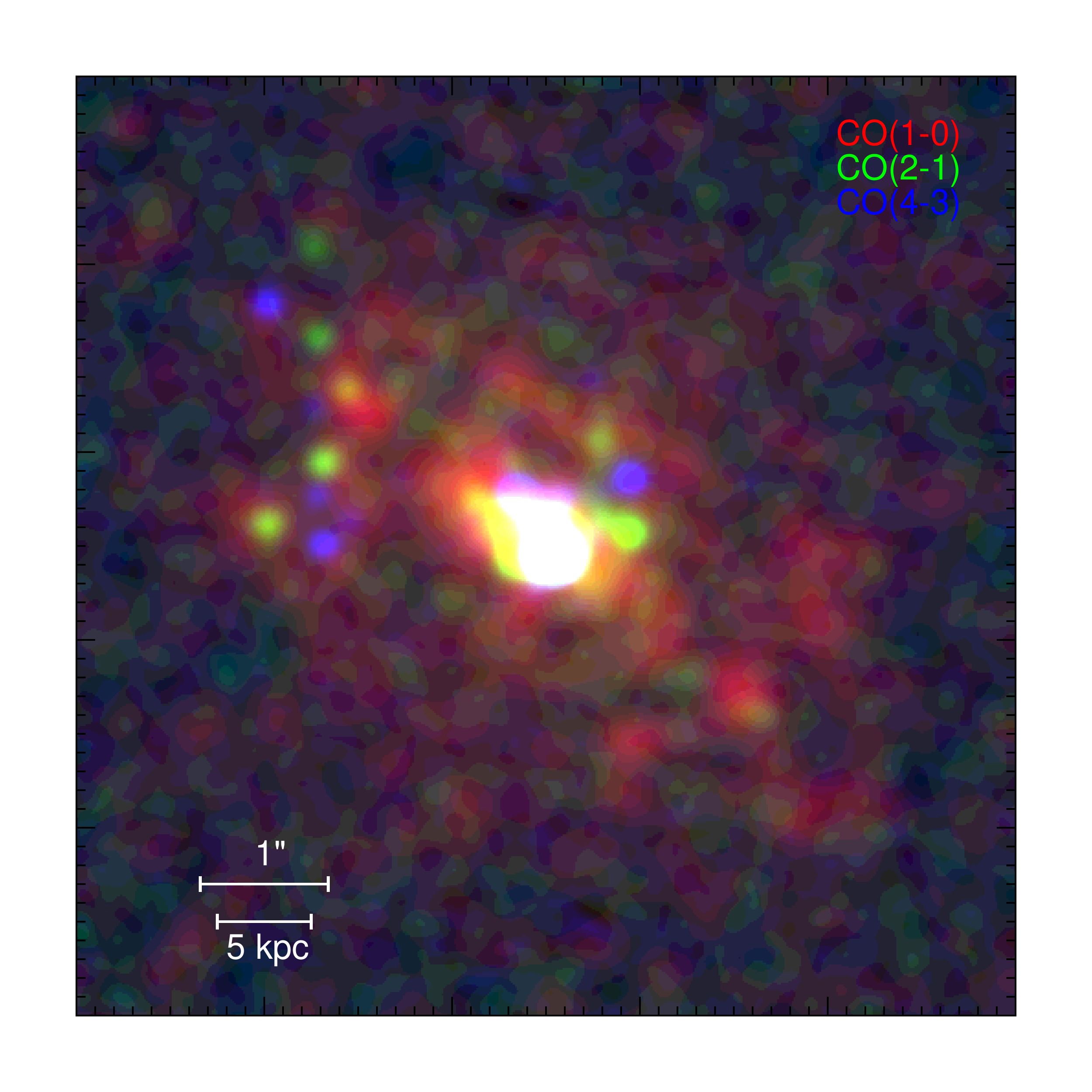}
\caption{At top: the ngVLA CO(1-0) moment 0 map overlaid with contours
  of CO(1-0), CO(2-1) (both from the ngVLA) and CO(4-3) emission (from
  ALMA).  The respective beamsizes, 0.''5 beam in CO(1-0) and 0.''3
  beam in both CO(2-1) and CO(4-3), are illustrated in the lower
  lefthand corner.
At bottom: side-by-side comparison of the raw CO emission in the
simulated galaxy against what would be observable with the ngVLA in
tricolor. Note that ALMA can observe CO(4-3), and the sole difference
is the expected required on-source time to reach a matched depth: $<$1
hour on source with ngVLA vs. $\sim$12 hours on source with current
ALMA.}
\label{fig5}
\end{figure}

The third and forth rows of Figure~\ref{fig4} show lower resolution
maps of the same kinematic structure, with a 0$''$.3 beam.  It is
reassuring that the underlying kinematics are intact at this low
resolution, and the same physical conclusions would be reached in
analyzing these maps as those with much higher spatial resolution:
that the system is dispersion dominated.  While the two views of this
system at different resolutions may provide consistent
interpretations, it is important to emphasize that work on
artificially redshifting $z\sim0$ high-SNR, high-resolution galaxy CO
data cubes to the same resolution and SNR as typical high-$z$ galaxies
has shown that it is easy to draw different conclusions as to the
physical drivers of a galaxy's evolution \citep{hung15a}.  Much like
the higher resolution images, the CO(2-1) fails to probe the total
underlying gas reservoir.

\section{Simulating Observables with CASA}

We pass the simulated position-position-velocity maps (with very high
spatial resolution and velocity resolution) through CASA to simulate
the imaging capabilities of the ngVLA.  Imaging for these
high-redshift galaxies is needed on $\sim$0$''$.1 scales and so does
not require the extremely long ngVLA baselines.  However, diffuse gas
emission on large scales is of general interest and so the more
compact baselines, in concert with the extended baselines on the
plateau, will be crucial to the recovery of good image quality.

Figure~\ref{fig5} shows a direct comparison between the observable
molecular gas reservoir in the three transitions and tri-color and the
underlying intrinsic gas distribution.  The mock ngVLA observations
mimic the simulated data, also showing the extended reservoir of
CO(1-0) beyond the core that is quite luminous in CO(2-1) and CO(4-3).
The ngVLA images also show detection of several knots of higher-J CO
emission, which trace smaller molecular gas complexes in the galaxy's
halo.

\begin{figure}
\centering
\includegraphics[width=0.75\textwidth]{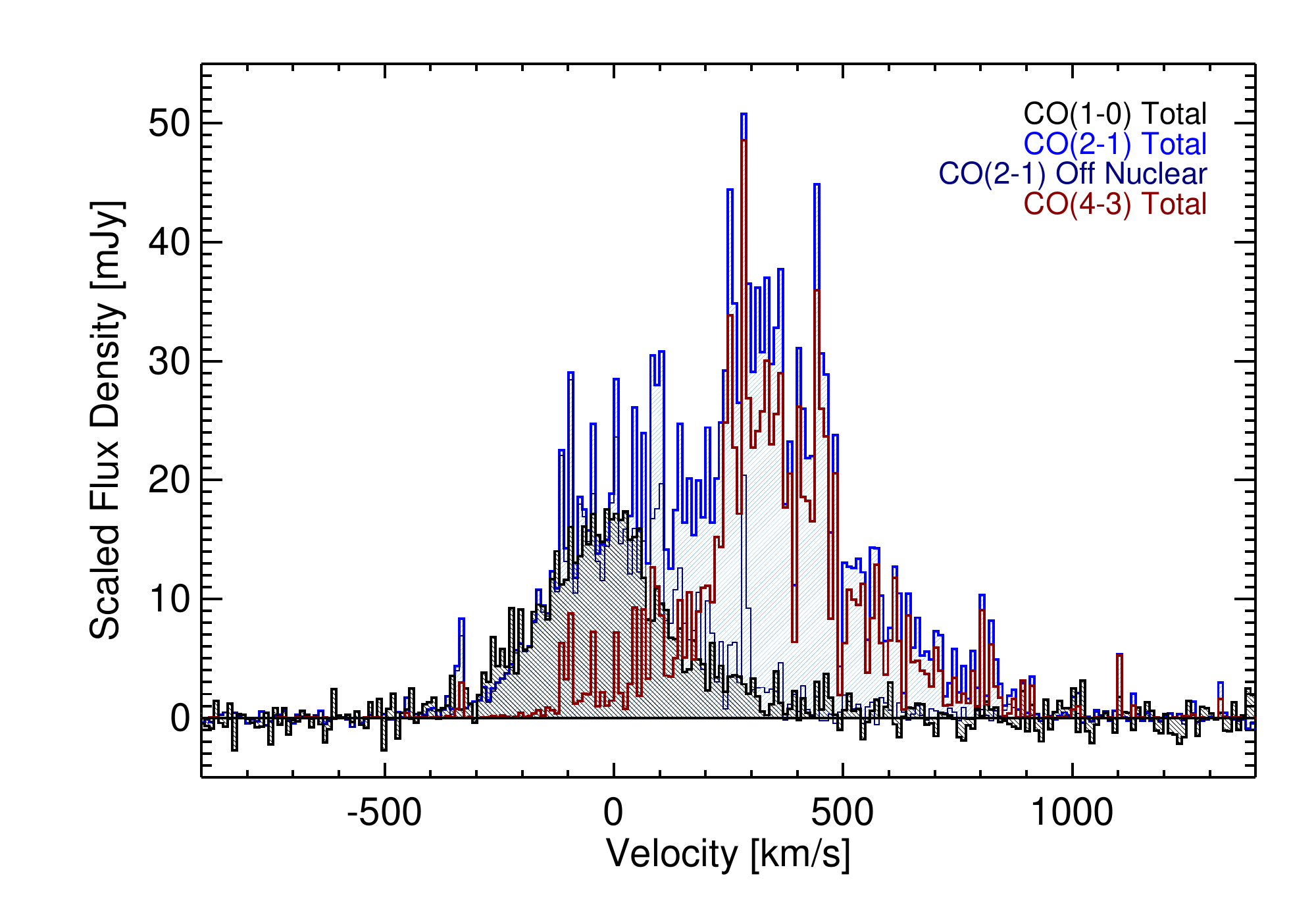}
\caption{The CO spectra of the simulated galaxies centered on the
  systemic redshift, which is well-probed by the peak line flux of
  CO(1-0) (black).  All spectra have been renormalized for clear
  viewing of the kinematic components of the gas reservoir.  CO(2-1)
  is double peaked, with an extended emission component coincident
  with the CO(1-0) emission and compact nuclear emission, which is
  both brighter and offset $\sim$400\,km\,s$^{-1}$ from the systemic
  velocity.  CO(4-3) emission primarily traces this nuclear offset
  peak with both low and high dispersion gas.}
\label{fig6}
\end{figure}

\section{Kinematically Distinct Gas Reservoirs}

In this particular case study, the moment zero line intensity maps
shown above already highlight fundamental differences between
different CO tracers in a high-z starbursting system.  However, these
differences are not isolated to the spatial extent of the gas.
Figure~\ref{fig6} shows the spectra for each of the three lines and
highlight how they are also kinematically distinct.  The ground-state
transition of CO, extended $\sim$13\,kpc along the semi-major axis of
the galaxy, is centered on 0\,km/s with a 300\,km/s FWHM.  CO(2-1)
exhibits a very different kinematic signature, comprised of two
components -- gas that is kinematically aligned with the CO(1-0)
emission and gas that is substantially offset at +350\,km/s, as well
as significantly brighter.  Higher-J lines, like the ALMA-observed
CO(4-3) traces only this offset component, which is perfectly aligned
with the galaxy's nucleus.

This has very important implications on deriving dynamical masses,
inferring kinematics of galaxies, and measuring gas masses, depletion
times, and star-formation efficiencies for galaxies at high-z, because
the vast majority of high-z galaxies are and will be characterized by
their high-J CO tracers, and not CO(1-0).  This demonstrates a key
need for the ngVLA: tracing the fundamental cold gas reservoir in
high-z galaxies as a core focus of extragalactic astrophysics.

\section*{Acknowledgments}

This work was completed with the generous support of the National
Radio Astronomy Observatory next generation VLA community students
program grant.  CMC, JBC, and CLH thank UT Austin for support through
the College of Natural Sciences.


\bibliography{caitlin-bibdesk}

\end{document}